\documentclass[preprint,12pt]{elsarticle}
\usepackage{graphicx}
\usepackage{dcolumn}
\usepackage{bm}
\usepackage{subfig}
\usepackage{color}
\usepackage{amsmath}
\usepackage{amssymb}
\begin{document}

\begin{frontmatter}

\title{Isotropic discrete Laplacian operators from lattice hydrodynamics}

\author[emu]{Sumesh P. Thampi}
\author[emu]{Santosh Ansumali}
\author[imsc]{Ronojoy Adhikari} 
\author[iac]{Sauro Succi} 
\address[emu]{Engineering Mechanics Unit, Jawaharlal Nehru Centre for Advanced Scientific Research, Bangalore 560064, India}
\address[imsc]{The Institute of Mathematical Sciences, CIT Campus, Chennai 600113, India}
\address[iac]{Istituto Applicazioni Calcolo, CNR Roma - via dei Taurini 9, 00185, Roma, Italy, EU}

\begin{abstract}
We show that discrete schemes developed for lattice hydrodynamics provide an elegant and physically
transparent way of deriving Laplacians with isotropic discretisation error. 
Isotropy is guaranteed whenever the Laplacian weights follow from the discrete Maxwell-Boltzmann equilibrium since these are, by construction, isotropic on the lattice. We also point out that stencils using as few as 15 points in three dimensions, generate isotropic Laplacians. 
These computationally efficient Laplacians can be used in cell-dynamical and hybrid lattice Boltzmann simulations, in favor of 
popular anisotropic Laplacians, which make use of larger stencils. 
The method can be extended to provide discretisations of higher order and for 
other differential operators, such the gradient, divergence and curl.
\end{abstract}

\end{frontmatter}

\section{Introduction}
Isotropy, an essential property of the Laplacian operator, is a desirable feature in any of its discrete representations. However, most commonly used discretisations of the Laplacian suffer from error terms that are anisotropic. 
These infect numerical solutions with anisotropies when, in fact, physical solutions are required to be isotropic. 
Two recent studies have addressed this issue, by providing algebraic methods for constructing isotropic Laplacians  \cite{Kumar2004, Patra2005}. 
In this paper, we show that lattices and associated weights used in lattice hydrodynamic simulations, naturally provide 
discrete Laplacians with isotropic discretisation error. 
These lattices and weights have been known for a long time in the lattice Boltzmann 
literature, \cite{Qian1992, SucciBook}, but  their connection to isotropic Laplacians has not been recognized. 
Here, we make this connection explicit and also show how discretisations of other differential 
operators like the gradient and divergence follow from it. 

Below, we first provide our main results for the isotropically discretised Laplacian. We then provide a derivation of our result and  make explicit its connection with the discrete Maxwell-Boltzmann equilibrium on the lattice. We conclude with a comparison of the isotropy properties of our Laplacians with those proposed earlier. 

\begin{figure}
\begin{center}
\includegraphics[width=0.7\linewidth]{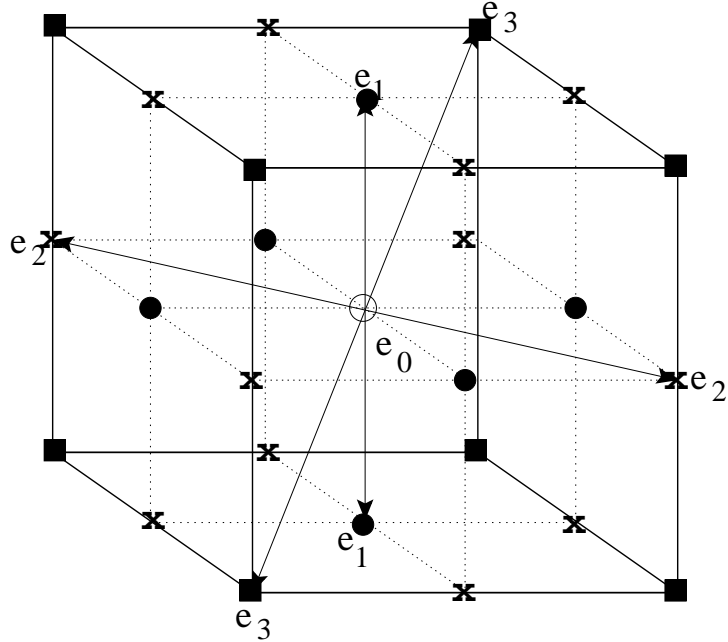}
\caption{\label{fig:lapcube}Ordering of points on a cubic unit cell according to ``energy shells'', as explained in the text. Here $e_0, e_1, e_2, e_3$ represent the energy shells corresponding to $e=0, 1, 2, 3$, marked by $\circ, \bullet, \boldsymbol{\times}$ and {\tiny $\blacksquare$} respectively. For clarity,  only one pair is highlighted.}
\end{center}
\end{figure}

Consider the cubic cell in Fig. \ref{fig:lapcube} with center $e_0$, the $6$ vectors ${\bf c}_i^1 $ that point to the face centers $e_1$, the $12$ vectors ${\bf c}_i^2$ that point to the edge centers $e_2$  and the $8$ vectors ${\bf c}_i^3$ that point to the vertices $e_3$. The face center vectors point along the Cartesian axes and so have the form $(\pm 1, 0, 0)$ with two vanishing Cartesian components, the edge center vectors are confined to the Cartesian planes and so have form $(\pm 1, \pm 1, 0)$ with one vanishing Cartesian component, while the vertex vectors have the form $(\pm 1, \pm 1, \pm 1)$. (With this choice, the discretisation step is set to unity.) This suggests an ``energy shell'' classification, where the energy is identified with the squared modulus of the vector. Then, the 27 points of the cubic cell lie on the ``energy shells'' with energies  $e_j = 0, 1, 2, 3$. There is 1 null vector ${\bf c}^0$  on the energy shell $e=0$, $6$ vectors ${\bf c}_i^1$ on the energy shell $e=1$, $12$ vectors ${\bf c}_i^2$ on the energy shell $e=2$ and $8$ vectors ${\bf c}_i^3$ on the energy shell $e=3$. In the cell dynamics literature these are known as the nearest neighbors - NN, next nearest neighbors - NNN, and next next nearest neighbors - NNNN, respectively.

For a function $\psi({\bf r})$ defined on such a cubic grid, we use the convenient shorthand $\psi({\bf r} + {\bf c}_i^j) = \psi_i^{(j)}$ for its value on the $i$-th point of the $j$-th energy shell. Then, our main results for the isotropic Laplacian $L({\bf r}) \equiv \nabla^2 \psi(\mathbf{r})$ are
\begin{align}
L(\mathbf{r})_{D2Q9} &= \frac{1}{6} \left[ 4 \sum_{i=1}^{4} \psi_i^{(1)} + \sum_{i=1}^{4} \psi_i^{(2)} - 20 \psi^{(0)} \right] \label{eqn:d2q9lap} \\
L(\mathbf{r})_{D3Q19} &= \frac{1}{6} \left[2 \sum_{i=1}^{6} \psi_i^{(1)} + \sum_{i=1}^{12} \psi_i^{(2)} - 24 \psi^{(0)} \right] \label{eqn:d3q19lap} \\
L(\mathbf{r})_{D3Q15} &= \frac{1}{12} \left[8 \sum_{i=1}^{6} \psi_i^{(1)} + \sum_{i=1}^{8} \psi_i^{(3)} - 56 \psi^{(0)} \right] \label{eqn:d3q15lap} \\
L(\mathbf{r})_{D3Q27} &= \frac{1}{36} \left[16 \sum_{i=1}^{6} \psi_i^{(1)} + 4 \sum_{i=1}^{12} \psi_i^{(2)}+\sum_{i=1}^{8} \psi_i^{(3)}- 152 \psi^{(0)} \right]
\label{eqn:d3q27lap}
\end{align} 
The first of these is a two-dimensional Laplacian which uses the $4$ face centers and $4$  edge centers on a Cartesian plane, while the remaining are all three dimensional Laplacians. The subscript on each of these Laplacians indicates that they derive from a D$n$Q$m$ lattice hydrodynamic model, as we explain in detail below. 

\section{Laplacians from lattice equilibria} 
The lattice formulation of kinetic theory provides a computationally efficient algorithm for solving the Navier-Stokes and related equations \cite{SucciBook}. A central quantity in lattice kinetic theory is the discrete form of the Maxwell-Boltzmann velocity distribution. There are several routes by which these equilibria can be obtained \cite{Maxwell1860, Boltzmann1872}. Here, we focus on the lattice generalization of the earliest derivation by Maxwell \cite{Maxwell1860}, which appeals only to factorisability and isotropy. Briefly, Maxwell argued that the distribution function $f({\bf c})$ must be isotropic in velocity space and so must be function of $c^2 = {\bf c}\cdot{\bf c}$ and it must be factorisable so that $f({\bf c})$ = $f({\bf c}_x)f({\bf c}_y)f({\bf c}_z)$. 
The only function that satisfies both these requirements is the Gaussian $f({\bf c}) \sim \exp(-{c^2/2 k_bT})$, whose variance 
$k_BT$ is fixed by requiring consistency with thermodynamics, and whose prefactor is fixed by normalization. In the above, the mass has been set to unity,
Recent works \cite{Ansumali2003, Karlin2010} show that a cubic lattice containing 27 velocities, made out of the direct product of the velocity set $\{-1, 0, 1\}$, has both the isotropy and factorisability features required by the continuum Maxwellian. 
This provides a succinct derivation of the lattice Maxwellian. The $27$ velocity lattice and the
weights so generated are called D$3$Q$27$ in the lattice hydrodynamics literature. 
Projections of this lattice to smaller velocity set are possible. In such projections, factorisability is lost, but isotropy is still maintained. 
It is well-known that isotropy of the lattice Maxwellians is a necessary condition for Navier-Stokes hydrodynamics on the lattice. 
It is this feature that we exploit to generate the isotropic Laplacians listed above. 

We begin with a D$n$Q$m$ lattice hydrodynamic model in $n$ dimensions with $m$ velocities. These have discrete weights $w_i^j$ which correspond to the energy shells $e_j$. For a scalar function $\psi({\bf r})$ defined on such a lattice, consider the transform
\begin{equation}
\langle \psi(\mathbf{r})\rangle = \sum_{j=0}^3 \sum_{i=1}^{N_j} w_{i}^j \psi(\mathbf{r}+\mathbf{c}_{i}^j)
\label{eqn:Phi2Psi}
\end{equation}
where $i$ labels the $i$-th discrete velocity in the $j$-th energy shell with $N_j$ velocities and $w_i^j$ is the corresponding weight factor. As is well known, the necessary conditions for obtaining isotropic Navier-Stokes hydrodynamics on the lattice are \cite{Hudong2008}
\begin{align}
\sum_{i,j} w_{i}^j &= 1 \label{eqn:ISO1} \\
\sum_{i,j} w_{i}^j c_{i,\alpha}^j c_{i,\beta}^j &= T \delta_{\alpha \beta} \label{eqn:ISO2} \\
\sum_{i,j} w_{i}^j c_{i,\alpha}^j c_{i,\beta}^j c_{i,\gamma}^j c_{i,\lambda}^j &= T^2 \Delta_{\alpha \beta \gamma \lambda}^{(4)}
\label{eqn:ISO3}
\end{align}
where Greek indices label Cartesian directions and $\Delta_{\alpha \beta \gamma \lambda}^{(4)} = \delta_{\alpha\beta} \delta _{\gamma\lambda} + \delta_{\alpha\lambda} \delta _{\gamma\beta} + \delta_{\alpha\gamma} \delta _{\beta\lambda}$. In the above $T$ is a lattice-dependent constant which identifies with the temperature of the moving particles. Using particles in the cubic cell with velocities $c_{i,\alpha}=\{-1,0,1\}$, the identity $c_{i,\alpha}^2=c_{i,\alpha}^4$ implies $T=1/3$ as the only value ensuring isotropy of the lattice hydrodynamics in the cubic cell. All weighted polynomials odd in the velocities vanish identically. In particular, linear, cubic and quintic polynomials are zero. On these lattices, the sextic polynomials are the first non-zero polynomials to break isotropy. 

Taylor expanding $\psi(\mathbf{r}+\mathbf{c}_{i}^j)$ in Eq. \ref{eqn:Phi2Psi} and applying the above symmetries of Eq. \ref{eqn:ISO1} - \ref{eqn:ISO3}, we obtain
\begin{equation}
\langle \psi(\mathbf{r}) \rangle = \psi(\mathbf{r}) + \frac{T}{2} \nabla^2 \psi (\mathbf{r}) + \frac{T^2}{8} \nabla^4 \psi (\mathbf{r}) + O(\nabla_{\alpha}^6).
\end{equation}
This equation can be solved for Laplacian $L({\bf r}) \equiv \nabla^2 \psi(\mathbf{r})$  to obtain
\begin{equation}
 L(\mathbf{r}) = \frac{2}{T} \left[\sum_{j=0}^3 \sum_{i=1}^{N_j} w_{i}^j \psi(\mathbf{r}+\mathbf{c}_{i}^j) - \psi(\mathbf{r})\right]+ O(\nabla^4).
\label{eqn:lapformula}
\end{equation}
This automatically secures isotropy of the Laplacian up to leading order error, with an error coefficient of order $O(T)$. 
This error cannot be made zero with the given cubic stencil. 
The above expression is remarkable, because any lattice with suitable weights which satisfies the conditions in Eq. \ref{eqn:ISO1} - \ref{eqn:ISO3} will provide an expression for the discrete Laplacian operator and will ensure isotropy. 
Writing down the terms of Eq. \ref{eqn:lapformula} explicitly, with  $\psi_i^{(j)}$ for $\psi(\mathbf{r}+\mathbf{c}_i^{j})$, we have
\begin{align}
 L(\mathbf{r}) = \frac{2}{T} \left[\sum_{i=1}^{6} w_{i}^1 \psi^{(1)} +\sum_{i=1}^{12} w_{i}^2 \psi^{(2)} +\sum_{i=1}^{8} w_{i}^3 \psi^{(3)} + (w^0-1) \psi^{(0)}\right].
\label{eqn:lapexp}
\end{align}
In other words, by redefining $\hat{w}^0 = w^0-1$ and $\hat{w}^j = w^j$ for $j=1,2,3$ we have
\begin{equation}
\sum_{i,j} \hat{w}_{i}^j = 0
\end{equation}
to replace Eq. \ref{eqn:ISO1} which, along with Eq. \ref{eqn:ISO2}-\ref{eqn:ISO3}, form a set of weights, $\{\hat{w}\}$, required to construct isotropic Laplacian operators in discrete space. These weights are lattice analogues of Hermite weights related to the Maxwell-Boltzmann equilibrium. Hence, this method of deriving isotropic Laplacians is an elegant and physically transparent way of calculating the operators compared to the methods present in the literature \cite{Kumar2004, Patra2005}.

\begin{table}
\begin{tabular}{|c|cc||c|c|c|c|c|}
\hline
  $e^j$ & $N_j$, &(for 2D) & $w_i^j$ &D2Q9 &D3Q15 & D3Q19 & D3Q27 \\
\hline
 0 & 1 &(1) &  $w_i^0$ & 4/9 &2/9 & 1/3 & 8/27 \\
\hline
  1 & 6 &(4) & $w_i^1$ & 1/9 &1/9 & 1/18 & 2/27 \\
\hline
  2 & 12 &(4) & $w_i^2$ &1/36 &0 & 1/36 & 1/54 \\
\hline
   3& 8 &(0) & $w_i^3$&0 &1/72 & 0 & 1/216 \\
\hline
\end{tabular}
\caption{Energy shells and the corresponding weight factors for various D$n$Q$m$ lattice hydrodynamics models. Values of $N_j$ for the two dimensional model D2Q9 are given in brackets.}
\label{tab:DnQm}
\end{table}

This general expression, when applied to the D$2$Q9, D$3$Q$15$, D$3$Q$19$, D$3$Q$27$ models, whose weights are listed in  table \ref{tab:DnQm}, gives the isotropic Laplacians we listed previously. The two-dimensional D2Q9 Laplacian 
was obtained earlier by a different argument in \cite{Kumar2004}.  Interestingly,  isotropic three-dimensional Laplacians can be achieved with just $15$ or $19$ velocities, i.e in D$3$Q$15$ and D$3$Q$19$ using Eq. \ref{eqn:lapformula}, and it is not necessary to use all $27$ velocities. This result was obtained earlier \cite{Gupta1998, Anant1987}, with weights identical to D3Q15 and D3Q19, without realizing 
the connection with lattice hydrodynamics. 

\section{Discussion} 
Having derived the Laplacians, we next proceed to compare their isotropy properties 
with other commonly used Laplacians, listed below.
\begin{align}
L(\mathbf{r})_{CD} &= \sum_{i=1}^{6} \psi_i^{(1)} - 6\psi^{(0)} \label{eqn:fdlap} \\
L(\mathbf{r})_{PK} &= \frac{1}{30} \left(14 \sum_{i=1}^{6} \psi_i^{(1)} + 3 \sum_{i=1}^{12} \psi_i^{(2)} + \sum_{i=1}^{8} \psi_i^{(3)} - 128 \psi^{(0)} \right) \label{eqn:pklap} \\
L(\mathbf{r})_{SO} &= \frac{1}{22} \left(6 \sum_{i=1}^{6} \psi_i^{(1)} + 3 \sum_{i=1}^{12} \psi_i^{(2)} + \sum_{i=1}^{8} \psi_i^{(3)} - 80 \psi^{(0)} \right)\label{eqn:solap} \\
L(\mathbf{r})_{KU} &= \frac{1}{48} \left(20 \sum_{i=1}^{6} \psi_i^{(1)} + 6 \sum_{i=1}^{12} \psi_i^{(2)} + \sum_{i=1}^{8} \psi_i^{(3)} - 200 \psi^{(0)}\right)\label{eqn:kulap} \\
L(\mathbf{r})_{EW} &= \frac{1}{9} \left( \sum_{i=1}^{6} \psi_i^{(1)} + \sum_{i=1}^{12} \psi_i^{(2)}+ \sum_{i=1}^{8} \psi_i^{(3)}- 26 \psi^{(0)} \right)
\label{eqn:lblap}
\end{align}
The suffixes $CD$,  $PK$, $SO$, $KU$,$EW$ stand for central difference, Patra-Kartunnen, Shinozaki-Oono, Kumar and `equally weighted' respectively. Eq. \ref{eqn:fdlap} is the standard central finite-difference expression. Eq. \ref{eqn:pklap} has been systematically derived by imposing conditions of rotational invariance and isotropy of the operator \cite{Patra2005, Spotz1996}. Eq. \ref{eqn:solap} is popular in the cell-dynamics and phase separation studies \cite{Shinozaki1993}. Eq. \ref{eqn:kulap} has been introduced as part of isotropic finite differences which describes discrete derivative operations without directional bias \cite{Kumar2004}. Eq. \ref{eqn:lblap} is a simple expression used in lattice Boltzmann simulations \cite{Kendon2001, Desplat2001} which gives equal weightage to all energy shells. In the small wave number limit, the discrete Fourier transform $L({\bf k}) = \sum_{{\bf r}}\exp(-i{\bf k\cdot r})L({\bf r})/\sum_{{\bf r}}\exp(-i{\bf k\cdot r})\psi({\bf r})$ of the Laplacian operator corresponding to the above expressions (Eq. \ref{eqn:d3q19lap} - \ref{eqn:d3q27lap}, \ref{eqn:fdlap} - \ref{eqn:lblap}) may be written as
\begin{align}
&L(\mathbf{k})_{D3Q15} = -k^2 + \frac{k^4}{12} + O(k_{\alpha}^6)\\
&L(\mathbf{k})_{D3Q19} = -k^2 + \frac{k^4}{12} + O(k_{\alpha}^6)\\
&L(\mathbf{k})_{D3Q27} = -k^2 + \frac{k^4}{12} + O(k_{\alpha}^6)\\
&L(\mathbf{k})_{CD} = -k^2 + \left[\frac{k^4}{12}-\frac{k_x^2 k_y^2 + k_x^2 k_z^2 + k_y^2 k_z^2}{6}\right] + O(k_{\alpha}^6)\\
&L(\mathbf{k})_{PK} = -k^2 + \frac{k^4}{12} + O(k_{\alpha}^6)\\
&L(\mathbf{k})_{SO} = -k^2 + \left[ \frac{k^4}{12} + \frac{k_x^2 k_y^2 + k_x^2 k_z^2 + k_y^2 k_z^2}{33/2}\right]+ O(k_{\alpha}^6)\\
&L(\mathbf{k})_{KU} = -k^2 + \frac{k^4}{12} + O(k_{\alpha}^6)\\
&L(\mathbf{k})_{EW} = -k^2 + \left[\frac{k^4}{12}+\frac{k_x^2 k_y^2 + k_x^2 k_z^2 + k_y^2 k_z^2}{6}\right] + O(k_{\alpha}^6)
\end{align}
\begin{figure*}
\subfloat[D3Q27, Eq. \ref{eqn:d3q15lap}]{\includegraphics[trim=45 50 10 45,clip, width=0.3\linewidth]{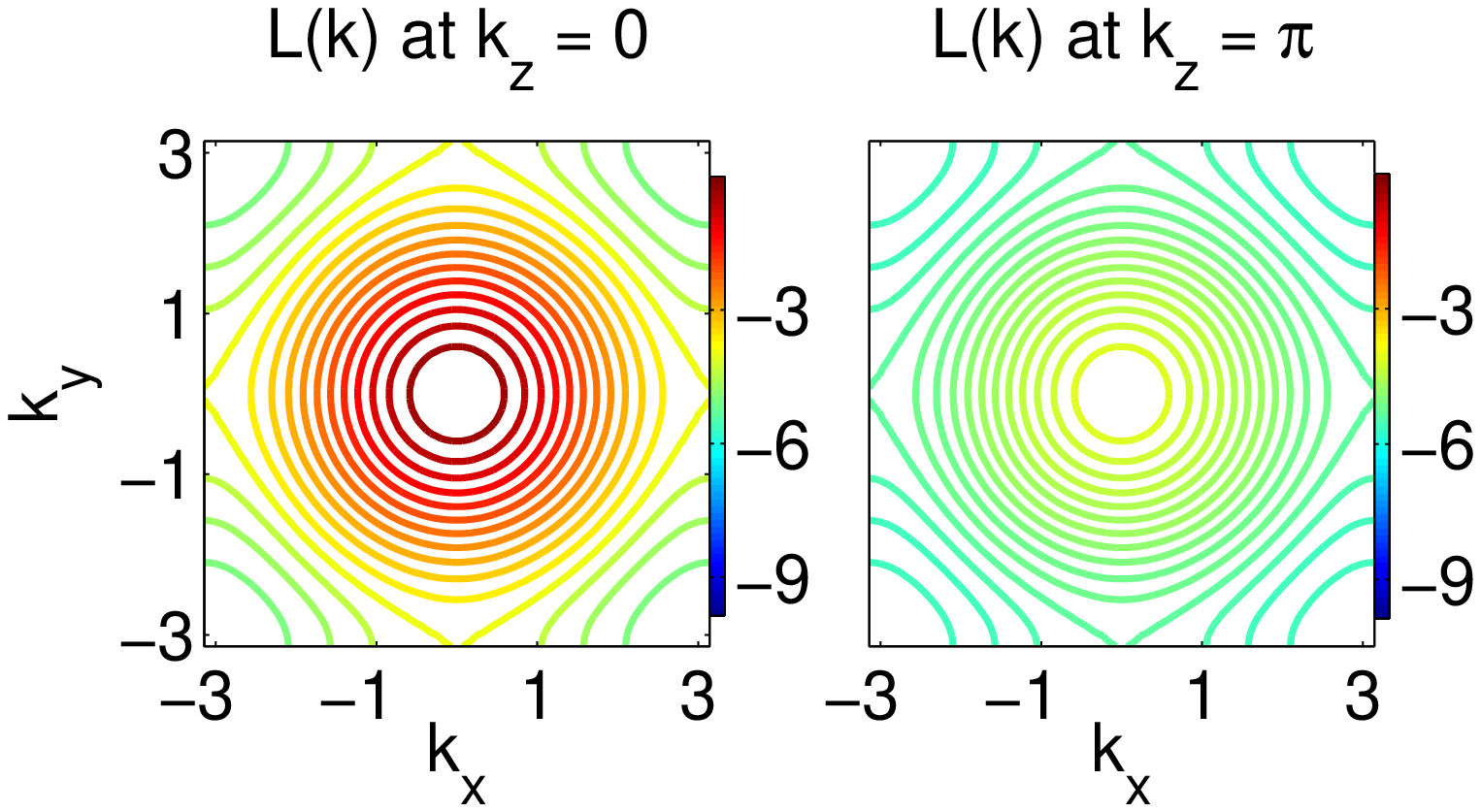}\label{fig:ft27}}
\hspace{4mm}
\subfloat[Patra-Kartunnen, Eq. \ref{eqn:pklap}]{\includegraphics[trim=45 50 10 45,clip, width=0.3\linewidth]{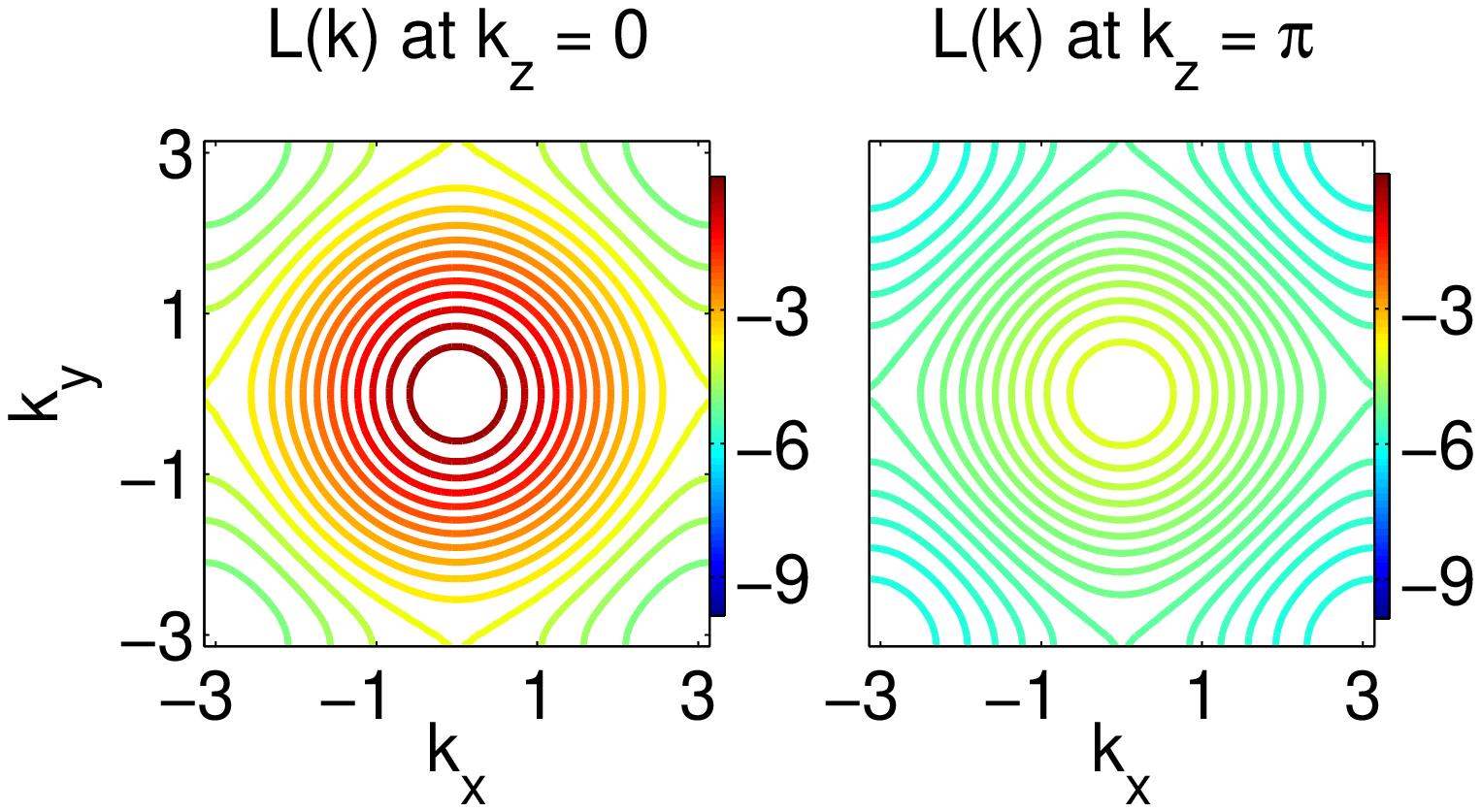}\label{fig:ftpk}}
\hspace{4mm}
\subfloat[Kumar, Eq. \ref{eqn:kulap}]{\includegraphics[trim=45 50 10 45,clip, width=0.3\linewidth]{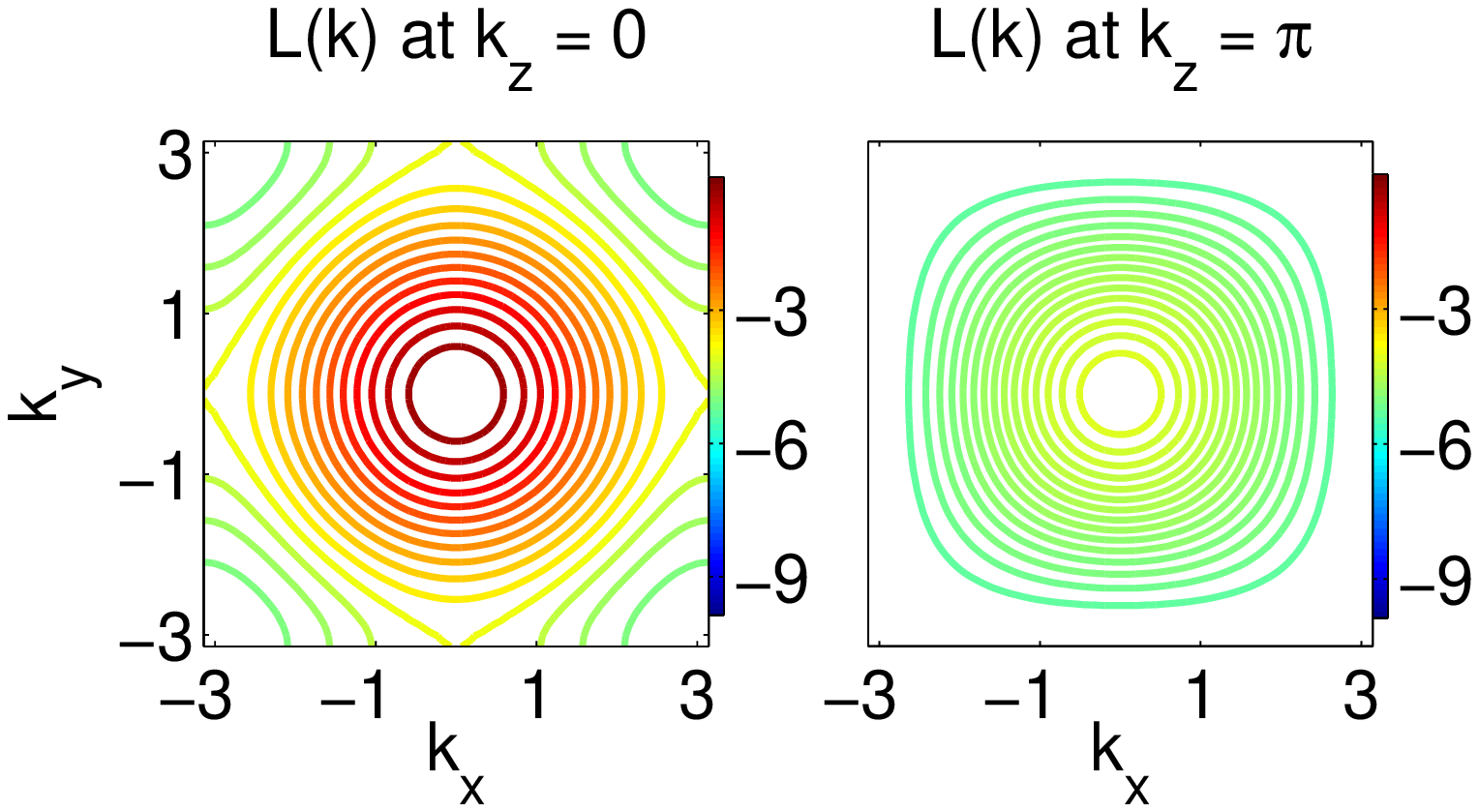}\label{fig:ftku}}\\
\subfloat[Shinozaki-Oono, Eq. \ref{eqn:solap}]{\includegraphics[trim=45 50 10 45,clip, width=0.3\linewidth]{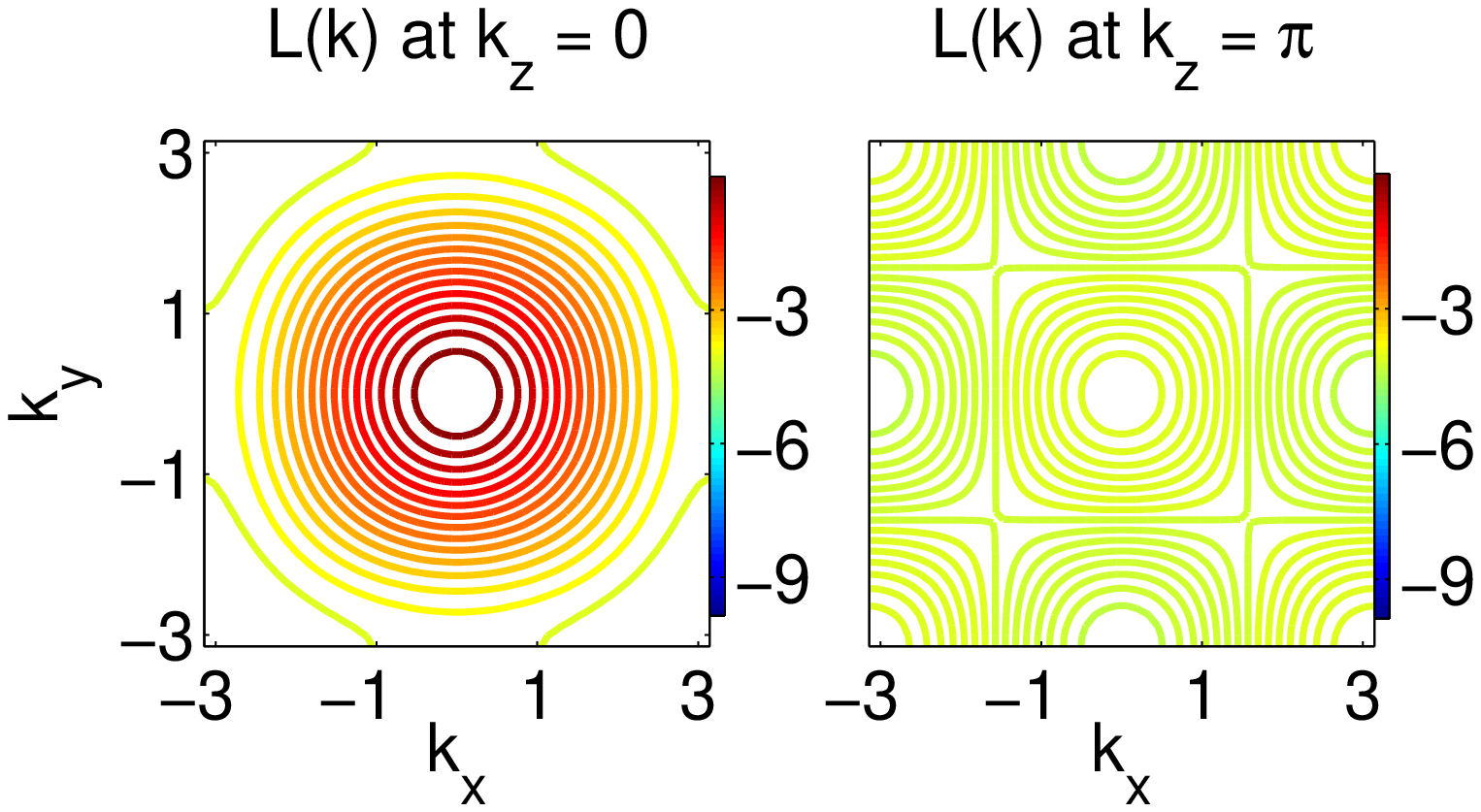}\label{fig:ftso}}
\hspace{4mm}
\subfloat[`Equally Weighted', Eq. \ref{eqn:lblap}]{\includegraphics[trim=45 50 10 45,clip, width=0.3\linewidth]{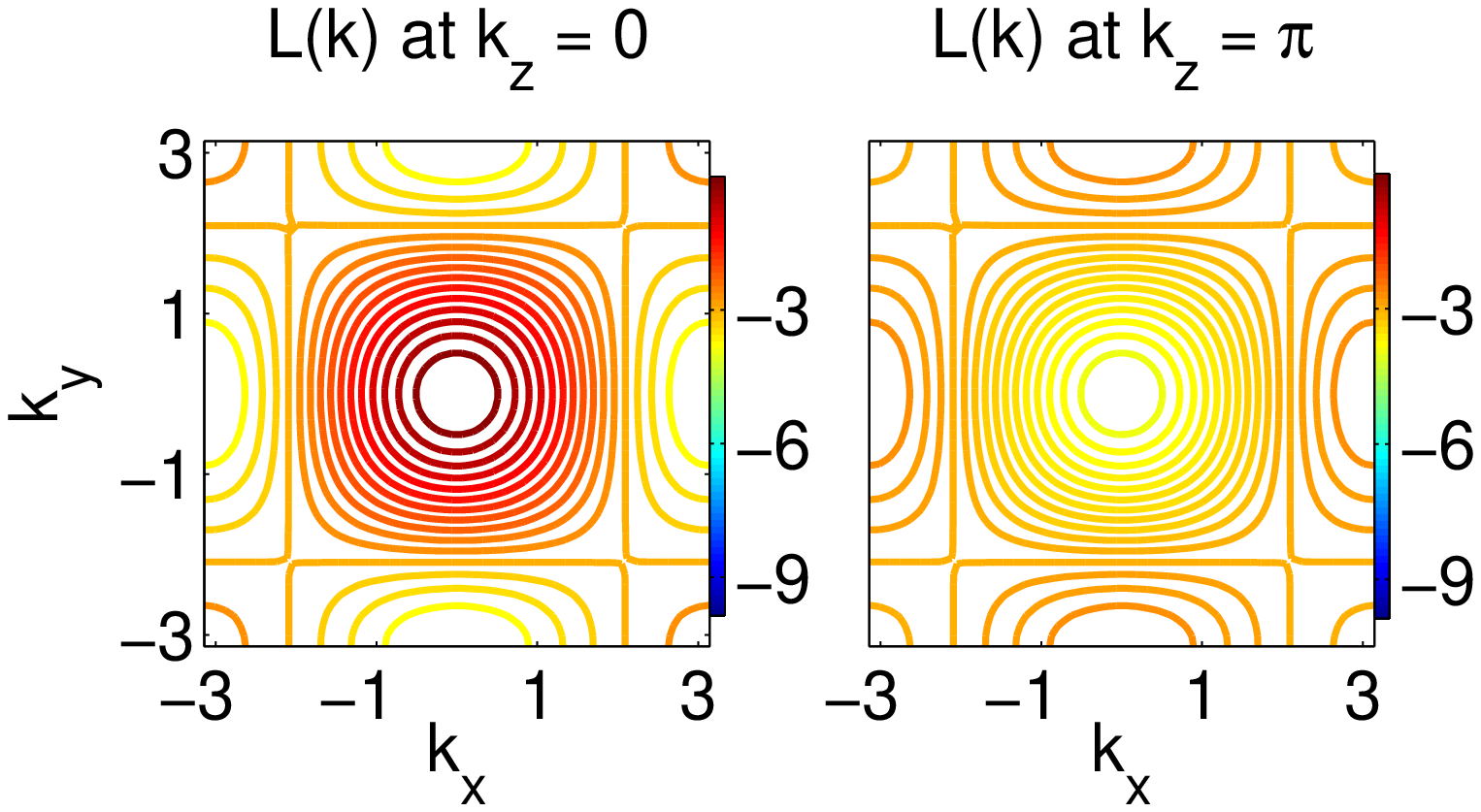}\label{fig:ftew}}
\hspace{4mm}
\subfloat[D3Q19, Eq. \ref{eqn:d3q19lap}]{\includegraphics[trim=45 50 10 45,clip, width=0.3\linewidth]{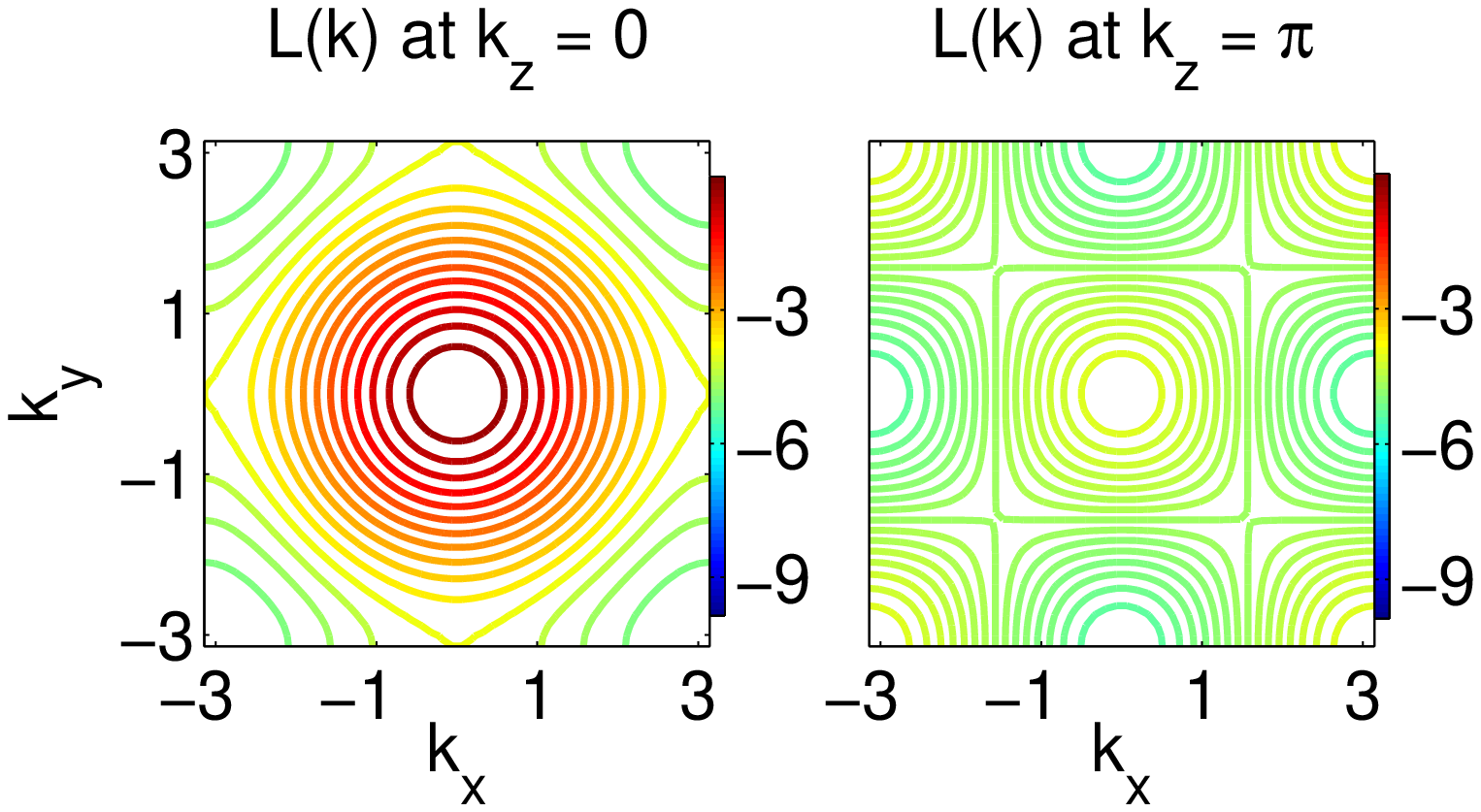}\label{fig:ft19}}\\
\subfloat[D3Q15, Eq. \ref{eqn:d3q15lap}]{\includegraphics[trim=45 50 10 45,clip, width=0.3\linewidth]{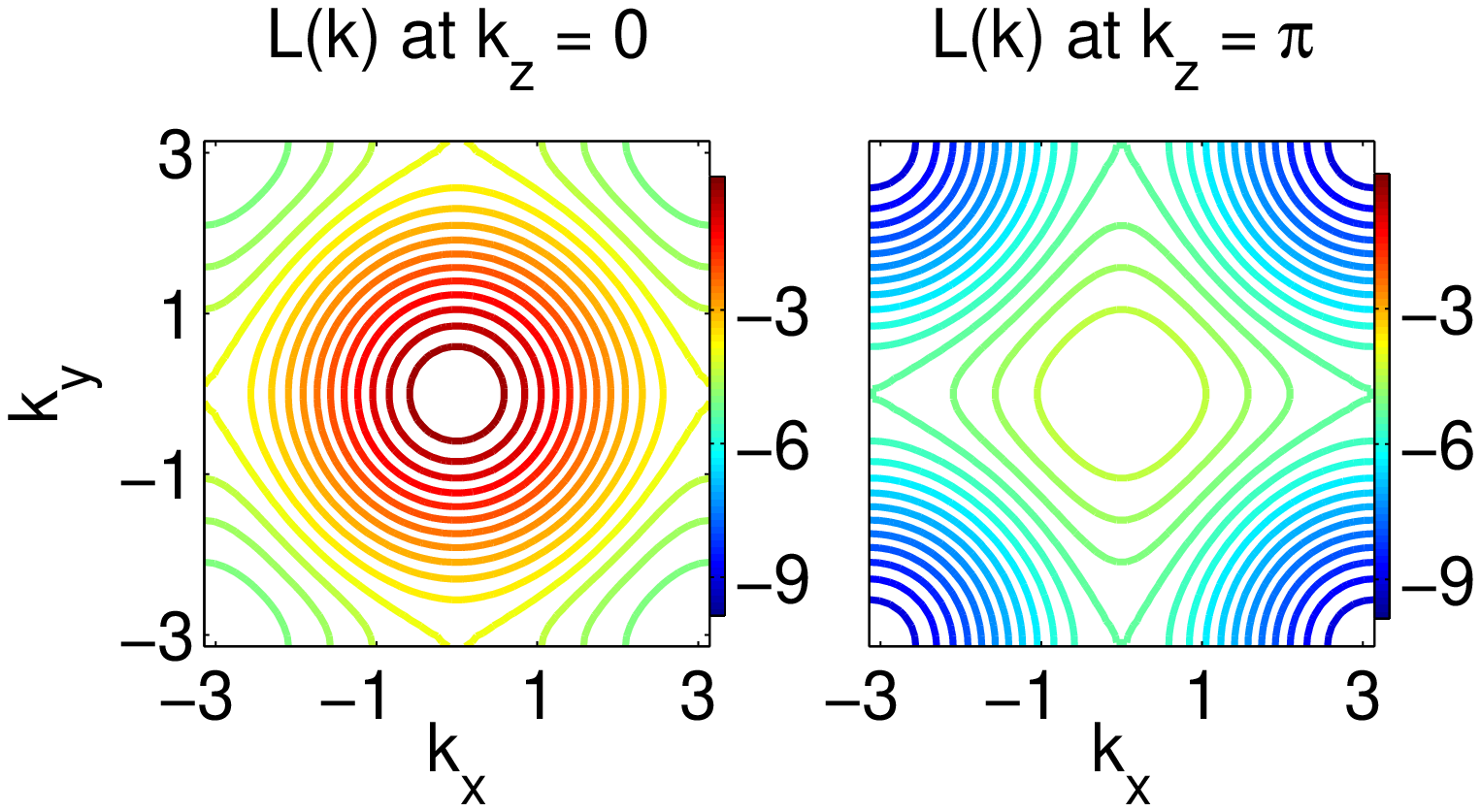}\label{fig:ft15}}
\hspace{4mm}
\subfloat[Central Difference, Eq. \ref{eqn:fdlap}]{\includegraphics[trim=45 50 10 45,clip, width=0.3\linewidth]{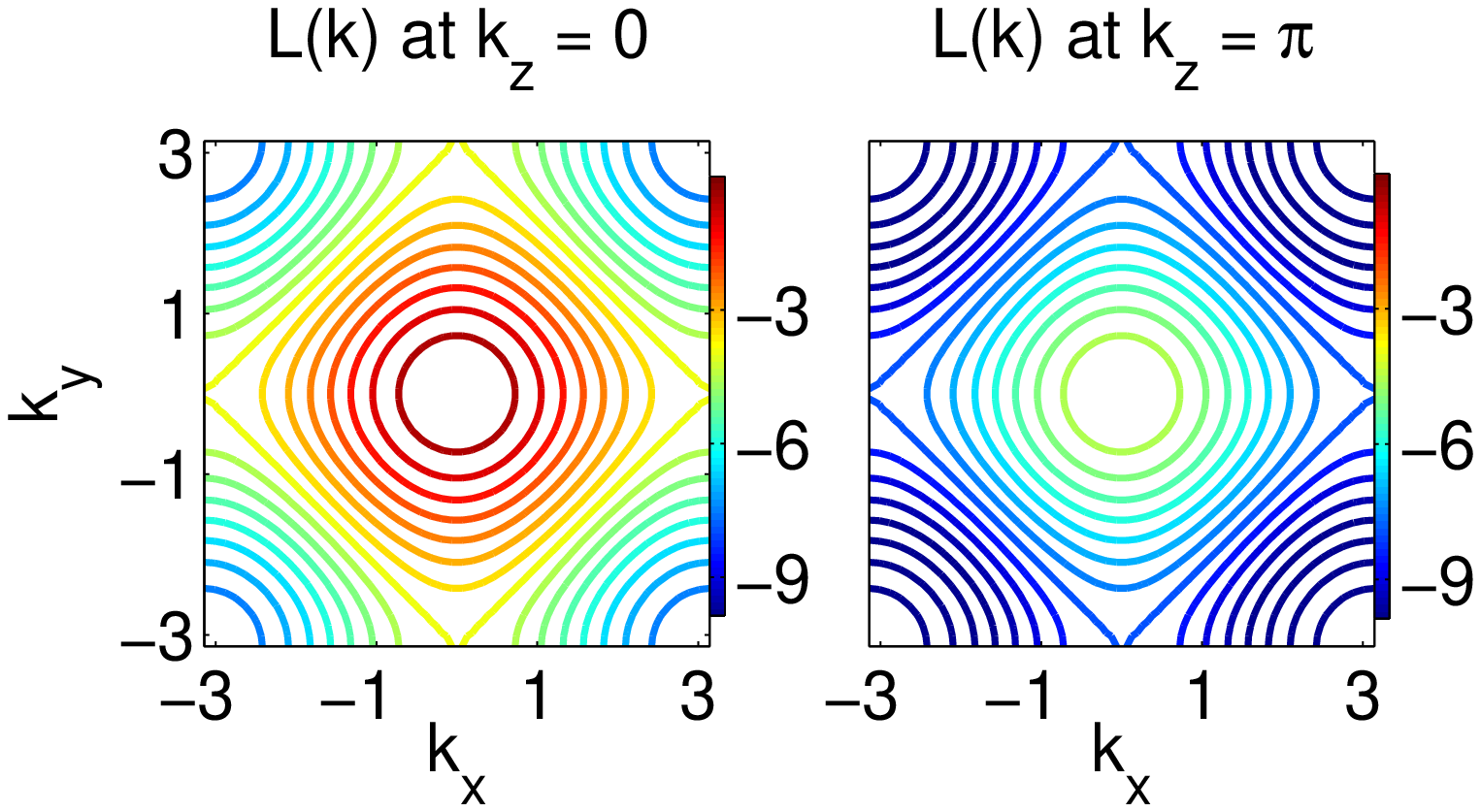}\label{fig:ftcd}}
\hspace{4mm}
\subfloat[$k^2$, Analytical expression]{\includegraphics[trim=45 50 10 45,clip, width=0.3\linewidth]{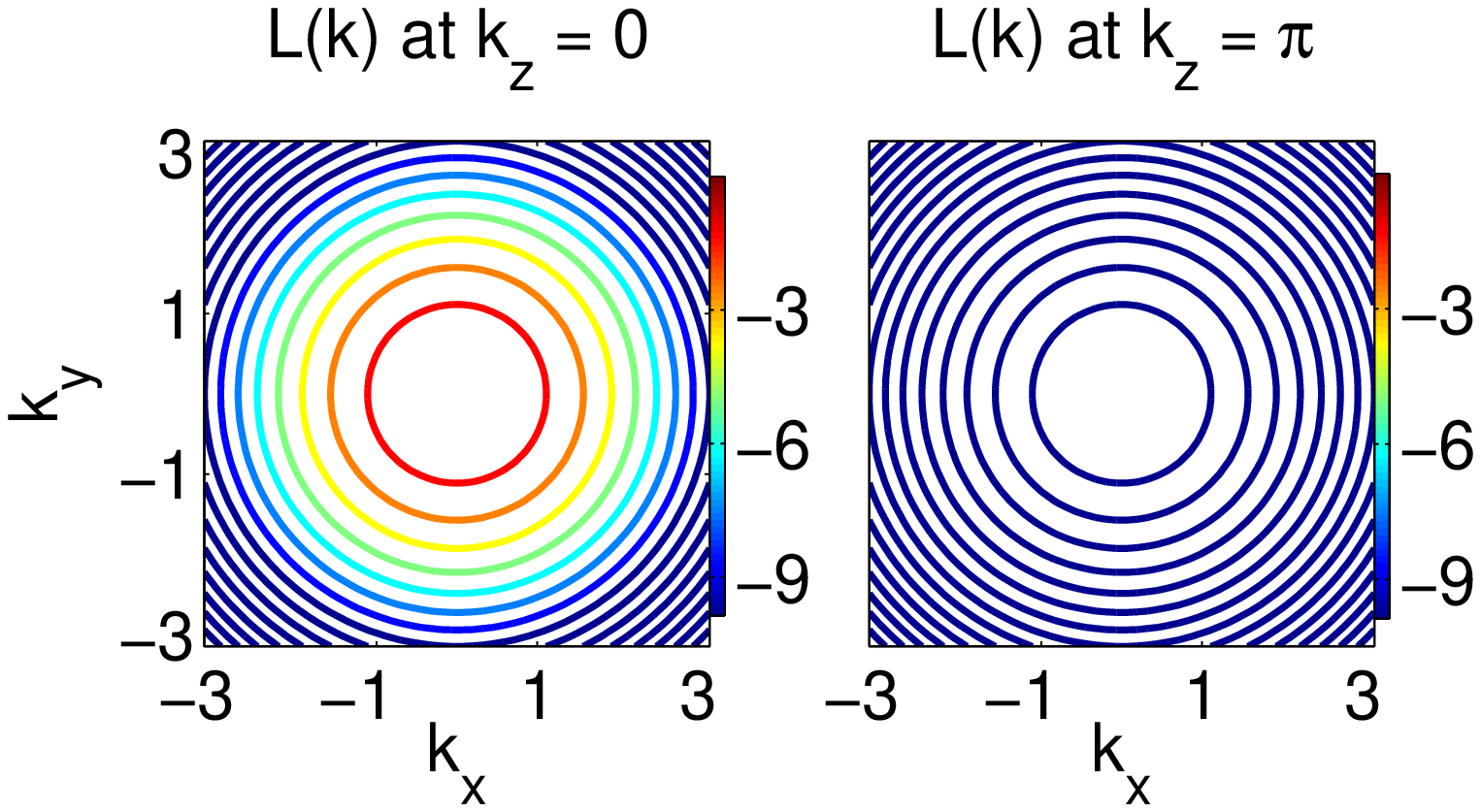}\label{fig:ftk2}}
\caption{Isocontours of the Laplacian operators in Fourier space, $L(\mathbf{k})$, at two different planes at $k_z = 0$ and $k_z = \pi$. Also shown is the isotropic plot of $k^2$ at these two planes for comparison. The color-bar is kept same for comparison across the operators. The y-axis is shown only for the first plot among each set for clarity.}
\label{fig:lapcont}
\end{figure*}
\begin{figure*}
\subfloat[D3Q27, Eq. \ref{eqn:d3q15lap}]{\includegraphics[trim=45 50 10 45,clip, width=0.3\linewidth]{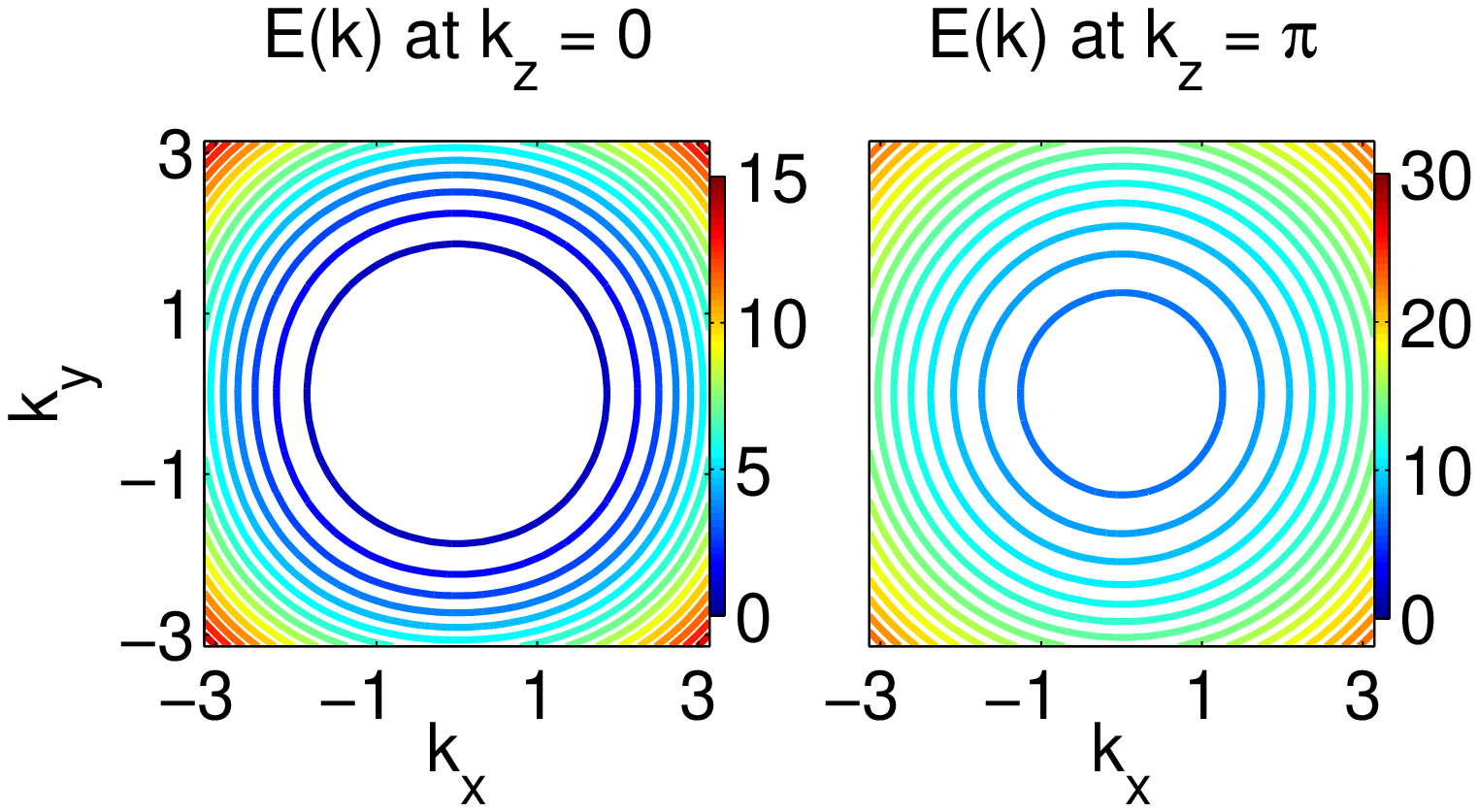}\label{fig:erft27}}
\hspace{4mm}
\subfloat[Patra-Kartunnen, Eq. \ref{eqn:pklap}]{\includegraphics[trim=45 50 10 45,clip, width=0.3\linewidth]{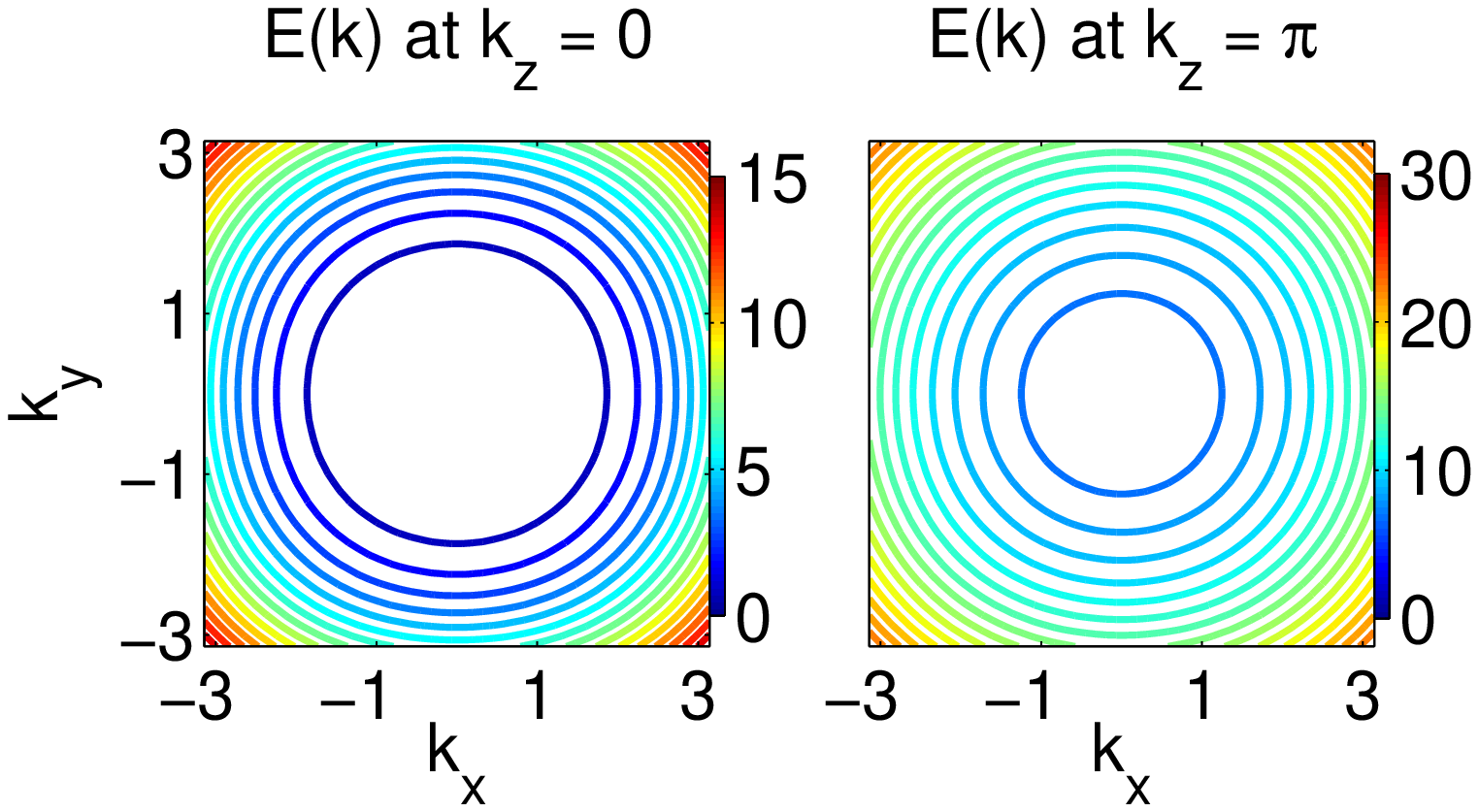}\label{fig:erftpk}}
\hspace{4mm}
\subfloat[Kumar, Eq. \ref{eqn:kulap}]{\includegraphics[trim=45 50 10 45,clip, width=0.3\linewidth]{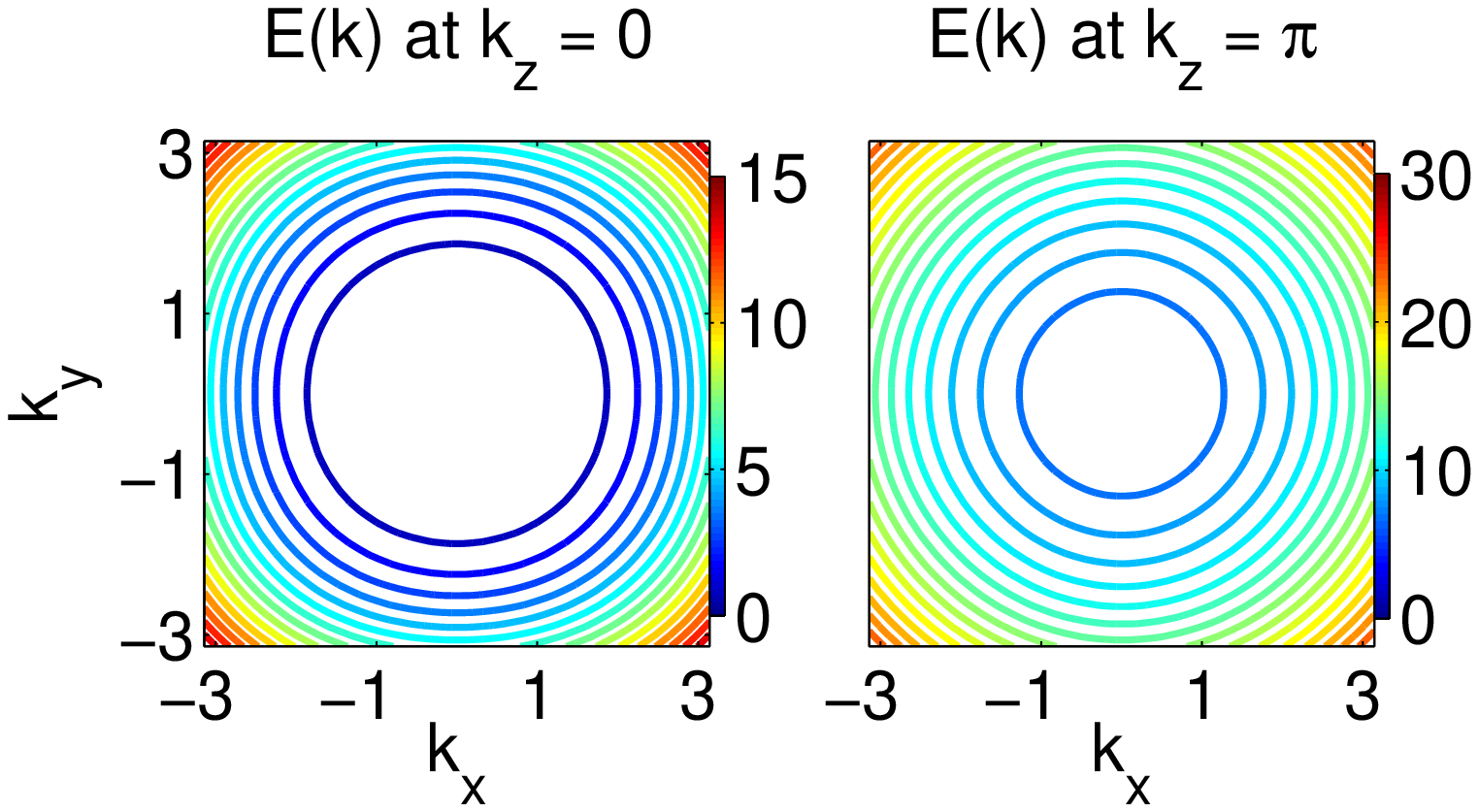}\label{fig:erftku}}\\
\subfloat[Shinozaki-Oono, Eq. \ref{eqn:solap}]{\includegraphics[trim=45 50 10 45,clip, width=0.3\linewidth]{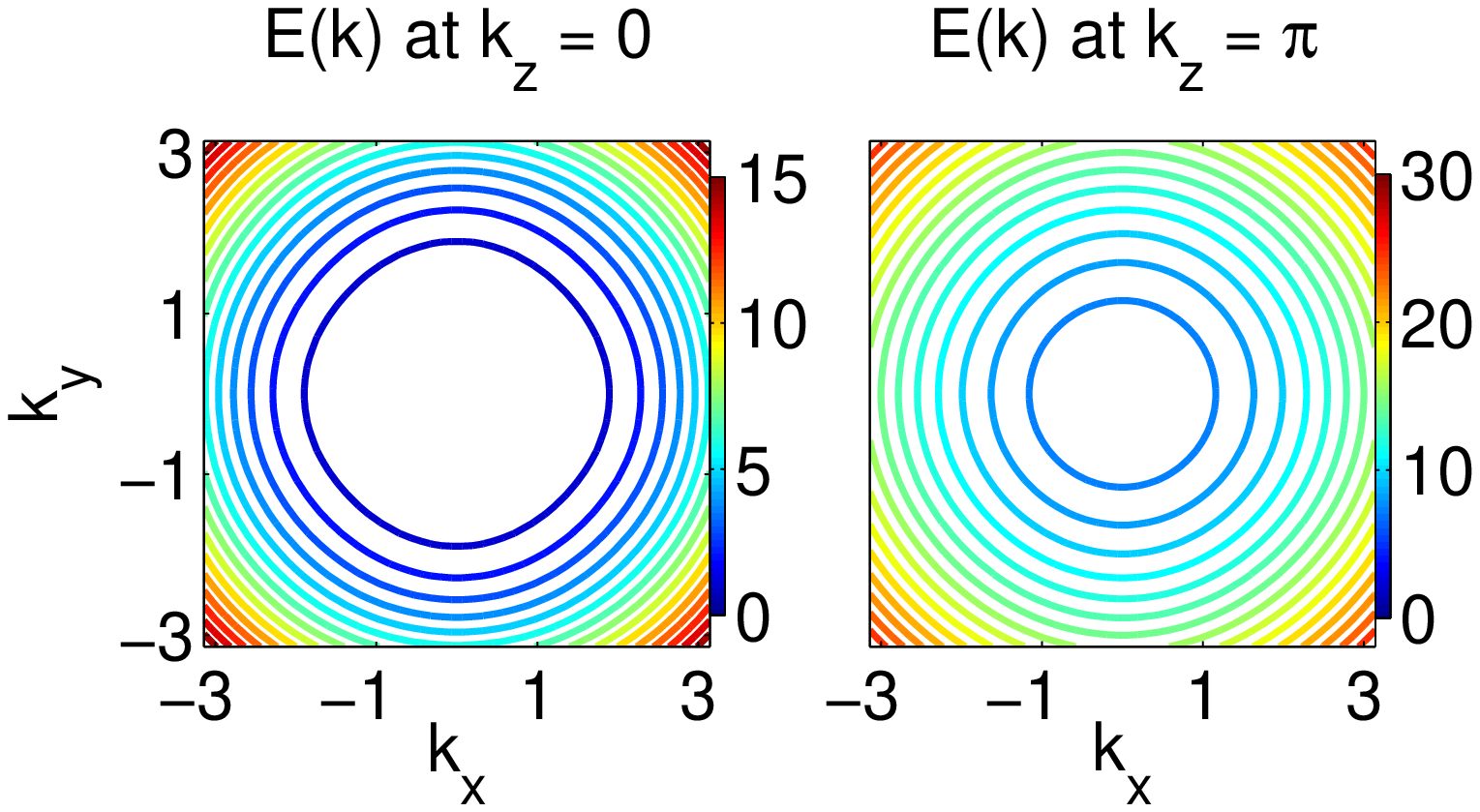}\label{fig:erftso}}
\hspace{4mm}
\subfloat[`Equally Weighted', Eq. \ref{eqn:lblap}]{\includegraphics[trim=45 50 10 45,clip, width=0.3\linewidth]{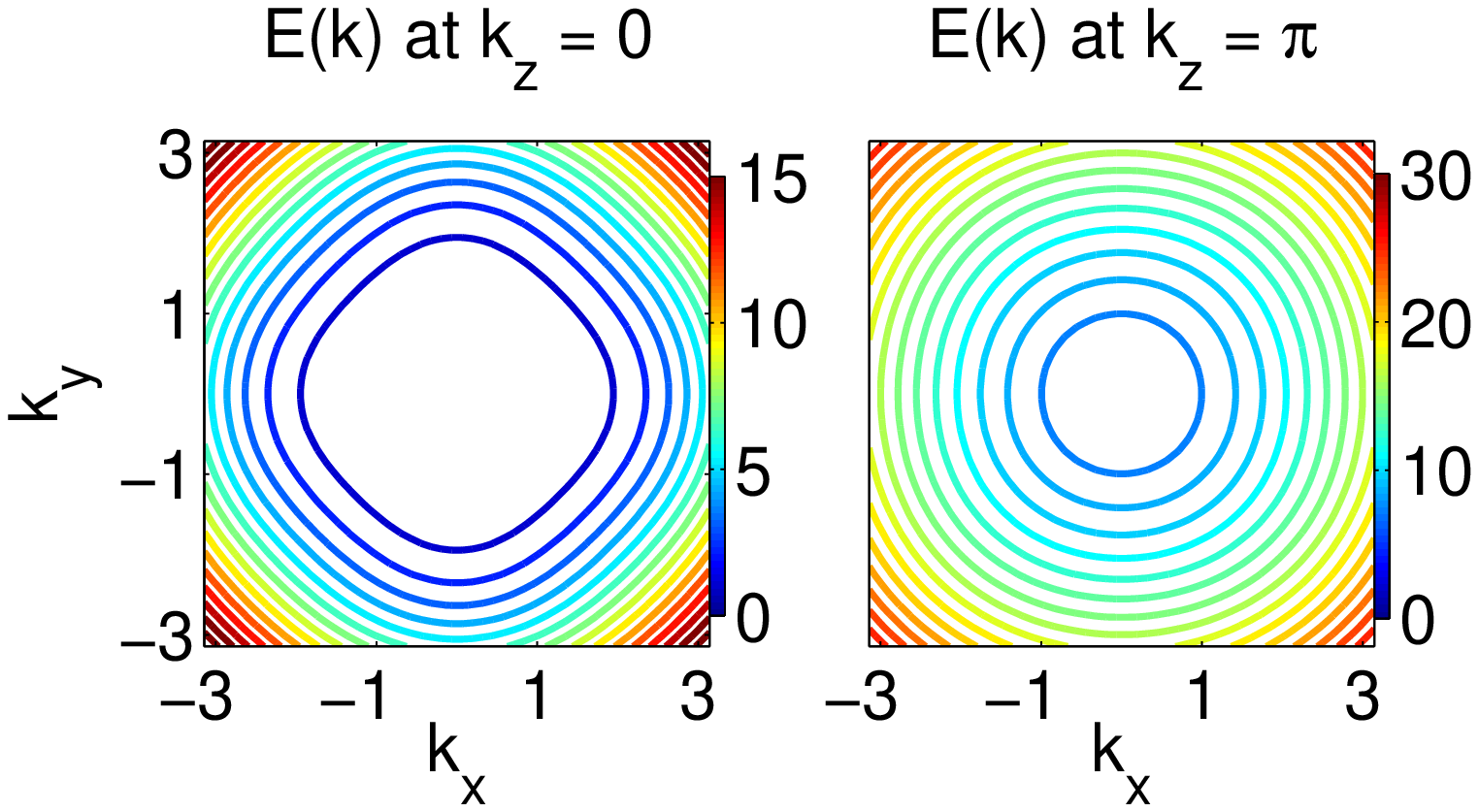}\label{fig:erftew}}
\hspace{4mm}
\subfloat[D3Q19, Eq. \ref{eqn:d3q19lap}]{\includegraphics[trim=45 50 10 45,clip, width=0.3\linewidth]{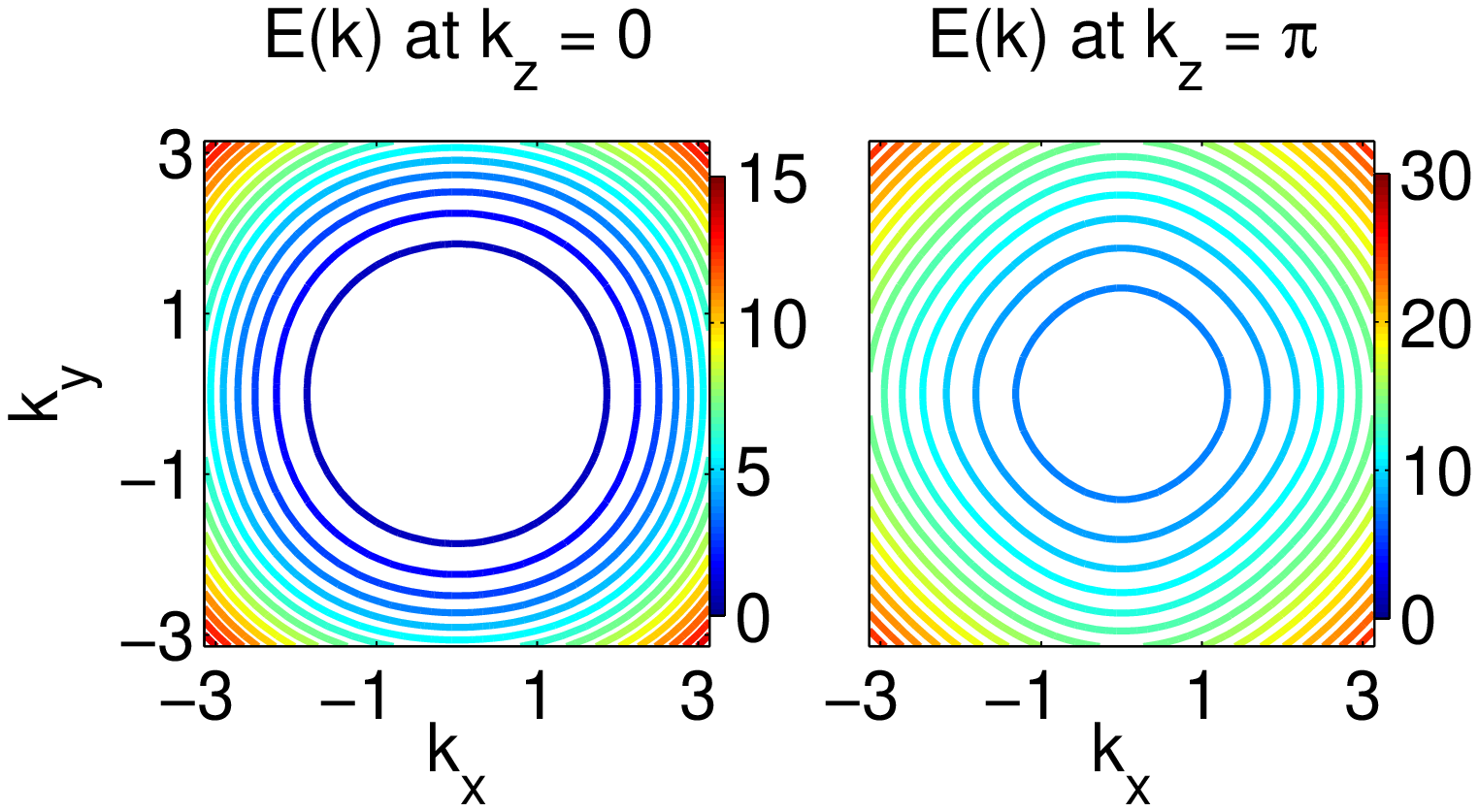}\label{fig:erft19}}\\
\subfloat[D3Q15, Eq. \ref{eqn:d3q15lap}]{\includegraphics[trim=45 50 10 45,clip, width=0.3\linewidth]{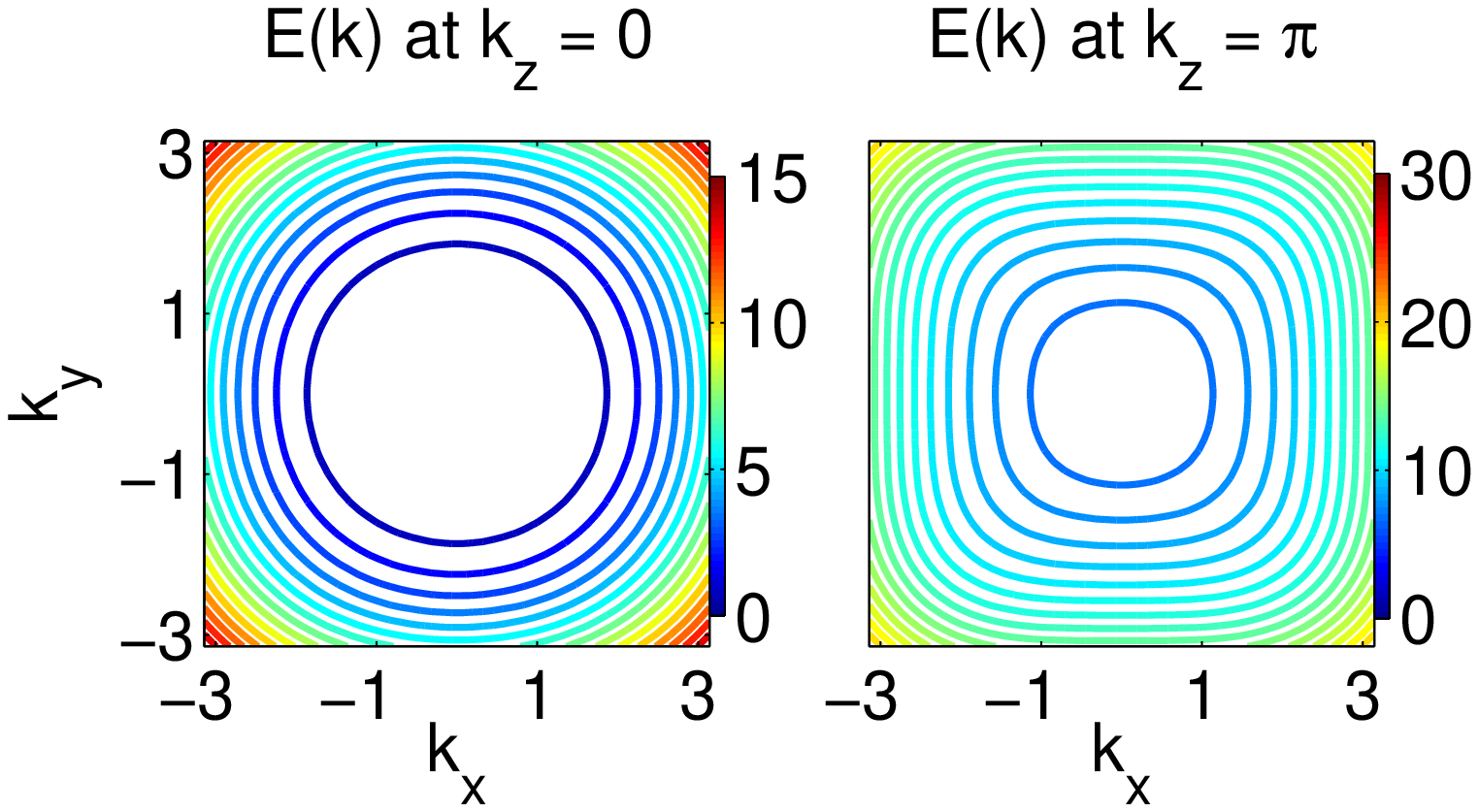}\label{fig:erft15}}
\hspace{4mm}
\subfloat[Central Difference, Eq. \ref{eqn:fdlap}]{\includegraphics[trim=45 50 10 45,clip, width=0.3\linewidth]{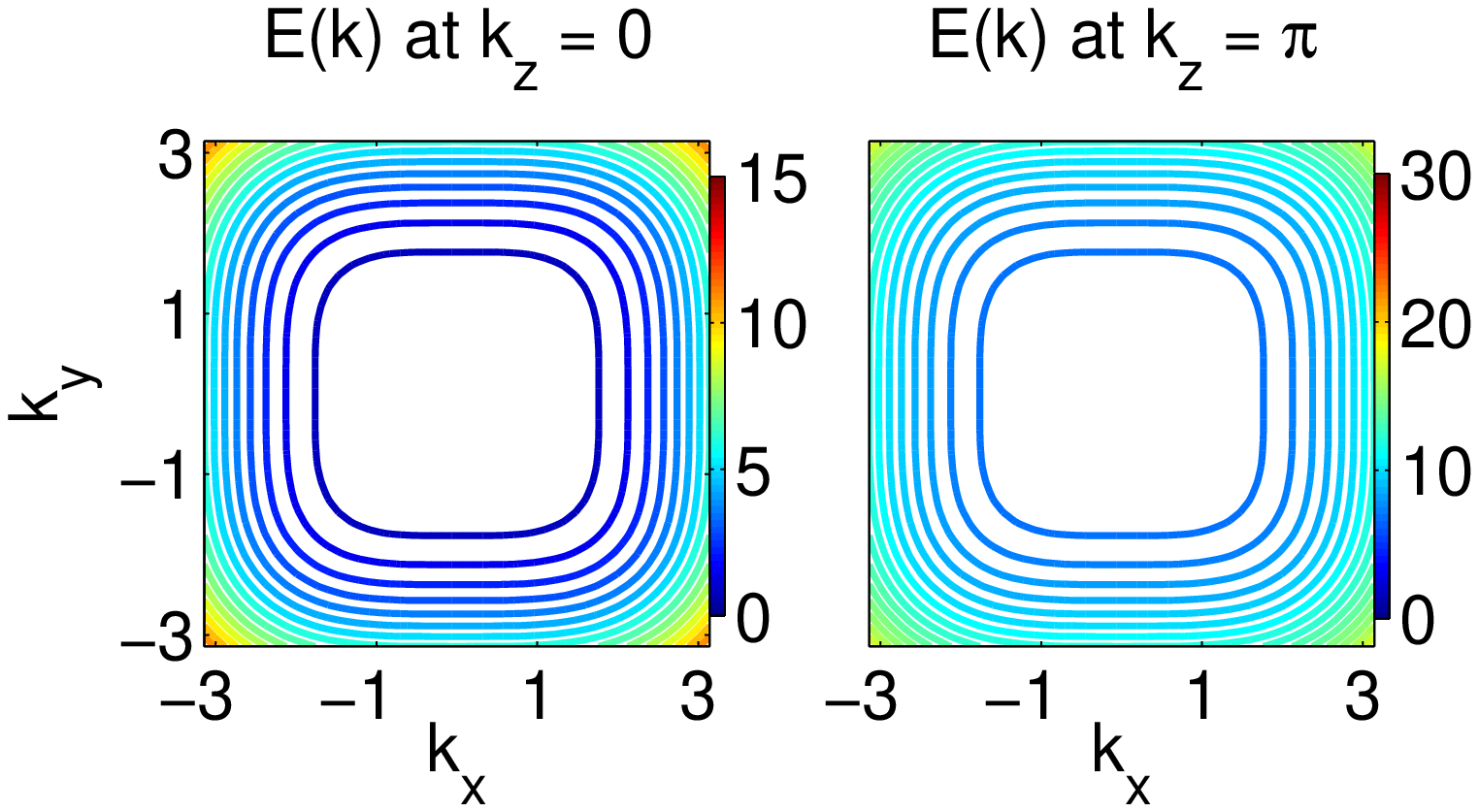}\label{fig:erftcd}}
\caption{Isocontours of the error of Laplacian operators in Fourier space defined as $E(\mathbf{k}) = |L(\mathbf{k})-(-\mathbf{k}^2)|$ at two different planes at $k_z = 0$ and $k_z = \pi$ (same as in Fig. \ref{fig:lapcont}). The color-bar is kept same for comparison across the operators. The y-axis is shown only for the first among each set for clarity.}
\label{fig:laperrs}
\end{figure*}
Isotropy at fourth order in $k$ is observed for all D$n$Q$m$ lattice stencils. 
This is also true for $PK$ and $KU$ stencils. Other stencils show anisotropic discretisation errors. 
While one may expect Eq. \ref{eqn:fdlap} to be anisotropic, due to the simplicity in construction, Eq. \ref{eqn:solap} and \ref{eqn:lblap} 
use all 27 points of the cubic cell, but are still not isotropic at leading order in error. 
None of these stencils provides isotropic error at sixth order. 

In Fig. \ref{fig:lapcont}, we plot isocontours of the Laplacian in Fourier space along the planes $k_z = 0, \pi$, to visually represent the degrees of anisotropy beyond quartic order. Both D$3$Q$27$ and Patra-Kartunnen Laplacians exhibit similar behavior and are the best of the set. The  Laplacian introduced by Kumar also exhibits comparable behavior. Both Shinozaki-Oono and `equally weighted' Laplacians are poor approximations at large wavenumbers and as seen from Eq. \ref{eqn:solap} - \ref{eqn:lblap} they are not isotropic at quartic order. Both D$3$Q$19$ and D$3$Q$15$ are isotropic at quartic order with smaller stencils consisting of $19$ and $15$ points. The smallest  of central difference stencil is the most anisotropic of the set, as seen in Fig. \ref{fig:ftcd}. Laplacians which have identical isotropic errors at quartic order may still have different stability properties when employed in numerical algorithms. This is part of ongoing work, and will be reported elsewhere. 
Our method naturally generalizes to other derivatives. For instance, the gradient of $\psi({\bf r})$ can be obtained from a Taylor expansion of the lattice transform $\sum_{ij} w_i^j {\bf c}_i^j \psi({\bf r} + {\bf c}_i^j)$. Following exactly the same steps used to derive the Laplacian, we see that discretisation error is isotropic. A systematic account of gradient, divergence and curl operators obtained by this method, as well as the case of genuine anisotropic physics \cite{rasin2005},  and higher order lattices \cite{sbragaglia2007,shan2008b}, will be reported elsewhere.

\section{Summary}

Summarizing, we have shown that lattices and weights commonly employed in lattice hydrodynamic simulations,  provide a computationally efficient discrete representation of the Laplacian preserving isotropy up to leading order error. The weights are derived from the lattice analogue of the Maxwell-Boltzmann distribution and are related to Hermite series expansion of local Maxwell-Boltzmann distributions. The use of these Laplacians should prove beneficial for cell dynamics simulations, hybrid lattice Boltzmann simulations, and many other problems, where efficient isotropic discretizations of the Laplacian are required.

 \section*{Acknowledgement}
One of the authors (SS) wishes to thank the Indian Academy of Sciences for financial support and kind hospitality.

\bibliographystyle{model1-num-names}
\bibliography{reference}

\end{document}